\documentclass[runningheads,orivec]{llncs}
\usepackage{amsmath,amssymb,amsfonts}
\usepackage{graphicx}
\usepackage{textcomp}
\usepackage{xcolor}
\usepackage{url}
\usepackage{multirow}
\usepackage{tikz}

\usepackage[noend]{algpseudocode}
\usepackage{algorithm}

\DeclareMathOperator*{\argmax}{arg\,max}

\usepackage{xcolor}

\allowdisplaybreaks

\usepackage[T1]{fontenc}
\usepackage{graphicx}

\usepackage[sectionbib,numbers]{natbib}

\begin{document}
\title{Slice and Explain: Logic-Based Explanations for Neural Networks through Domain Slicing}
\titlerunning{Slice and Explain}
%Carlos H. L. Cavalcante\inst{1}\orcidID{0000-0001-9395-8338
\author{Luiz Fernando Paulino Queiroz\orcidID{0009-0003-2407-3204} \and
Carlos Henrique Leitão Cavalcante\orcidID{0000-0001-9395-8338} \and 
Thiago Alves Rocha\orcidID{0000-0001-7037-9683}%\thanks{Corresponding author}
}
%Second Author\inst{1}\orcidID{1111-2222-3333-4444}}
% \and Third Author\inst{3}\orcidID{2222--3333-4444-5555}
%\inst{2,3}
\authorrunning{Queiroz et al.}
\institute{Instituto Federal do Ceará (IFCE), Maracanaú, Ceará, Brazil\\
\email{\{luiz.fernando, henriqueleitao, thiago.alves\}@ifce.edu.br}}

\maketitle              % typeset the header of the contribution
\begin{abstract}
Neural networks (NNs) are pervasive across various domains but often lack interpretability. To address the growing need for explanations, logic-based approaches have been proposed to explain predictions made by NNs, offering correctness guarantees. However, scalability remains a concern in these methods. This paper proposes an approach leveraging domain slicing to facilitate explanation generation for NNs. By reducing the complexity of logical constraints through slicing, we decrease explanation time by up to 40\% less time, as indicated through comparative experiments. Our findings highlight the efficacy of domain slicing in enhancing explanation efficiency for NNs.
\end{abstract}

\section{Introduction}
Neural Networks (NNs) are widely used across various domains \cite{goodfellow:16, alghoul2018email, musleh2019predicting}, but their lack of interpretability makes them opaque models \cite{koh2017understanding}. This means that the processes leading to their outputs from a given set of inputs are not inherently interpretable or explainable. This opacity raises concerns in critical systems with limited tolerance for failure \citep{Ribeiro2016Lime}. As a result, efforts have been made to develop tools that can explain NN behavior \cite{ignatiev2019abduction,elboher2020abstraction}. In this work, an explanation for a given input and its corresponding output is a subset of attributes that are sufficient to determine the output. While the remaining attributes are not necessary. For instance, consider an input
$\{income=5000$, $employment\_status=employed$, $loan\_history=good$, $age=30\}$ and let $N$ be a neural network that outputs \textit{credit\_approved} for this input. A possible explanation for this prediction is $\{loan\_history=good,employment\_status=employed\}$. This means that if an applicant has a good loan history and is employed, the neural network $N$ predicts \textit{credit\_approved}, regardless of the values of $income$ and $age$.
%weng2018evaluating
%Elboher
% audemard2024computation
%marques2020explaining
%Therefore, there is a need for alternative methods that ensure correctness.

Heuristic explainable AI (XAI) methods, such as LIME \cite{Ribeiro2016Lime} and Anchors \cite{ribeiro2018anchors}, have been used to provide explanations for NNs and other machine learning (ML) models. However, these methods lack guarantees of correctness  \cite{amparore2021trust}, making it difficult to assess the trustworthiness of machine learning (ML) models. Recent research in logic-based XAI for ML classifiers \citep{shi2018symbolic, ignatiev2019abduction, wang2021probabilistic, audemard2022preferred, rocha2025generalizing, bassan2023towards,bjorner2023formal} offers promising solutions by providing guarantees of correctness. Moreover, these methods also ensure explanations without redundancy \cite{ignatiev2019abduction}. Such explanations provide only essential information and seem easier to understand. Among these logic-based XAI approaches, the method proposed by \citet{ignatiev2019abduction} specifically targets NNs. Their approach encodes the NN as a set of logical constraints—comprising Boolean combinations of linear (in)equalities over binary and continuous variables—and leverages a Mixed Integer Linear Programming (MILP) solver to determine whether a subset of input features is sufficient to determine the same prediction. Nonetheless, scalability remains a key challenge when applying these techniques to larger NNs.

%However, logic-based explainability encounter scalability challenges when applied to large NNs.%rochafilho2024generalizing

%To improve the scalability of logic-based XAI, we explore a slicing-based approach inspired by NN simplification techniques~\cite{katz}. 
%not only enhances scalability but also preserves correctness and irredundancy guarantees of logic-based XAI. 
%n this work, we propose an enhancement of the method by Ignatiev et al., 
%a novel approach to improve the scalability of logic-based XAI by exploring the concept of \emph{slicing}

In this work, we propose an enhancement of the method by \citet{ignatiev2019abduction}, introducing a novel technique based on \emph{slicing} to improve scalability. Inspired by NN simplification techniques~\cite{katz}, this technique partitions the input domain into smaller subdomains, allowing us to simplify the original NN and potentially reduce explanation computation time. To the best of our knowledge, this is the first application of slicing in the context of XAI. By adapting this technique to explainability, we show that domain slicing can enhance the scalability of logic-based XAI while preserving correctness and irredundancy guarantees. Our experiments evaluate the impact of slicing on explanation times, considering configurations with up to 3 slices.

Our results show that, for deeper NNs, slicing can significantly reduce explanation times by over 40\% for some datasets. This suggests that slicing is a promising strategy for improving the scalability of XAI methods, particularly for larger NNs. However, the impact of slicing varies across datasets, and its effectiveness seems to depend on the characteristics of the data. For smaller NNs, the overhead of handling multiple slices often outweighed the potential gains in efficiency. These findings suggest that slicing has the potential to improve scalability for larger NNs in certain scenarios.

%In our experiments, we evaluated the impact of slicing input attributes on explanation times, considering configurations with up to 3 slices.
%The results suggest that slicing can reduce explanation times in specific cases, with some datasets showing improvements exceeding 40\%.

%However, the impact of slicing varies across datasets, and its effectiveness depends on the characteristics of the data. These findings indicate that slicing has the potential to improve scalability in certain scenarios.

%Results indicate that slicing can significantly reduce explanation times, with improvements exceeding 40\% in some cases. These findings highlight the potential of slicing to improve scalability in logic-based explainability.

\section{Preliminaries}

This section aims to present basic concepts necessary for a general understanding of the work.

\subsection{First-order Logic over LRA}\label{subsec:logic}

% Reviewer 3 (4) 
In this work, we use first-order logic (FOL) to generate explanations with guarantees of correctness. We use quantifier-free first-order formulas over the theory of linear real arithmetic (LRA) \cite{kroening2016decision}. Variables are allowed to take values from the real numbers $\mathbb{R}$. We consider formulas as $\sum^n_{i=1} w_i x_i \leq b$, such that $n \geq 1$ is a natural number, each $w_i$ and $b$ are fixed real numbers, and each $x_i$ is a first-order variable. As usual, if $F$ and $G$ are formulas, then $(F \wedge G), (F \vee G), (\neg F), (F \to G)$ are formulas. We allow the use of other letters for variables instead of $x_i$, such as $s_i$, $z_i$.
For example, $(2.5x_1 + 3.1s_2 \geq 6) \wedge (x_1=1 \vee x_1=2) \wedge (x_1=2 \to s_2 \leq 1.1)$ is a formula by this definition. Observe that we allow standard abbreviations as $\neg (2.5x_1 + 3.1s_2 < 6)$ for $2.5x_1 + 3.1s_2 \geq 6$.

% defined below:
% \begin{equation}
%         \begin{aligned}
%              F, G &:= p \mid ,\\
%              p &:= \sum^n_{i=1} w_i x_i \leq b, %\mid \sum^n_{i=1} w_i x_i < b,
%         \end{aligned}    
% \end{equation}
% such that $F$ and $G$ are quantifier-free first-order formulas over the theory of linear real arithmetic. Moreover, $p$ represents the atomic formulas  

Since we are assuming the semantics of formulas over the domain of real numbers, an \textit{assignment} $\mathcal{A}$ for a formula $F$ is a mapping from the first-order variables of $F$ to elements in the domain of real numbers. For instance, $\{x_1 \mapsto 2.3, x_2 \mapsto 1\}$ is an assignment for $(2.5x_1 + 3.1x_2 \geq 6) \wedge (x_1=1 \vee x_1=2) \wedge (x_1=2 \to x_2 \leq 1.1)$. An assignment $\mathcal{A}$ \textit{satisfies} a formula $F$ if $F$ is true under this assignment. For example, $\{x_1 \mapsto 2, x_2 \mapsto 1.05\}$ satisfies the formula in the above example, whereas $\{x_1 \mapsto 2.3, x_2 \mapsto 1\}$ does not satisfy it. Moreover, an assignment $\mathcal{A}$ \textit{satisfies} a set $\Gamma$ of formulas if all formulas in $\Gamma$ are true under $\mathcal{A}$.

A set of formulas $\Gamma$ is \textit{satisfiable} if there exists a satisfying assignment for $\Gamma$. To give an example, the set $\{(2.5x_1 + 3.1x_2 \geq 6), (x_1=1 \vee x_1=2), (x_1=2 \to x_2 \leq 1.1)\}$ is satisfiable since $\{x_1 \mapsto 2, x_2 \mapsto 1.05\}$ satisfies it. As another example, the set $\{(x_1 \geq 2), (x_1 < 1)\}$ is unsatisfiable since no assignment satisfies it. Given a set of formulas $\Gamma$ and a formula $G$, the notation $\Gamma \models G$ is used to denote \textit{logical consequence} or \textit{entailment}, i.e., each assignment that satisfies $\Gamma$ also satisfies $G$. As an illustrative example, let $\Gamma = \{x_1 = 2 , x_2 \geq 1\}$ and $G = (2.5x_1 + x_2 \geq 5) \wedge (x_1=1 \vee x_1=2)$. Then, $\Gamma \models G$. The essence of entailment lies in ensuring the correctness of the conclusion $G$ based on the set of premises $\Gamma$. In the context of computing explanations, as presented in \cite{ignatiev2019abduction}, logical consequence serves as a fundamental tool for guaranteeing the correctness of predictions made by NNs.

The relationship between satisfiability and entailment is a fundamental aspect of logic. It is widely known that, for all sets of formulas $\Gamma$ and all formulas $G$, it holds that
\footnotesize
\begin{equation}\label{ent_unsat}
\Gamma \models G \text{ if and only if } \Gamma \cup \{\neg G\} \text{ is unsatisfiable.}    
\end{equation}
\normalsize
%\[\]

\noindent For instance, $\{x_1 = 2, x_2 \geq 1),  \neg((2.5x_1 + x_2 \geq 5) \wedge (x_1=1 \vee x_1=2))\}$ has no satisfying assignment since an assignment that satisfies $\{x_1 = 2 , x_2 \geq 1\}$ also satisfies $(2.5x_1 + x_2 \geq 5) \wedge (x_1=1 \vee x_1=2)$ and, therefore, does not satisfy $\neg((2.5x_1 + x_2 \geq 5) \wedge (x_1=1 \vee x_1=2))$. Since our approach builds upon the concept of logical consequence, we can leverage this connection in the context of computing explanations for NNs.

% \begin{equation}\label{ent_unsat}
    
% \end{equation}

%\[\]

\subsection{Classification Problems and Neural Networks}

In machine learning, classification problems are defined over a set of $n$ input features $\mathcal{F} = \{x_1, ..., x_n\}$ and a set of $k$ classes $\mathcal{K} = \{c_1, c_2,...,c_k\}$. In this work, we consider that each input feature $x_i \in \mathcal{F}$ takes its values $v_i$ from the domain of real numbers. Moreover, each input feature $x_i$ has an upper bound $ub_i$ and a lower bound $lb_i$ such that $lb_i \leq x_i \leq ub_i$, i.e., its domain is the closed interval $[l_i, u_i]$. This is represented as an input domain $\mathcal{D} = \{l_1 \leq x_1 \leq u_1,\text{ }l_2 \leq x_2 \leq u_2, ..., l_n \leq x_n \leq u_n \}$. For example, a feature for the height of a person belongs to the real numbers and may have lower and upper bounds of $0.5$ and $2.1$ meters, respectively. As another example, a pixel intensity feature in an image classification problem belongs to the real numbers and typically has lower and upper bounds of $0$ and $255$, respectively, representing the range of grayscale values. Finally, a set $\mathcal{I} = \{x_1 = v_1, x_2 = v_2, ..., x_n = v_n\}$, such that each $v_i$ is in the domain of $x_i$, represents an instance of the input domain.

In this work, a NN is a function that maps elements in the input domain into the set of classes $\mathcal{K}$. Then, a NN is a function $\mathcal{N}$ such that, for all instances $\mathcal{I}$, $\mathcal{N}(\mathcal{I}) = c \in \mathcal{K}$. A NN is represented as $L+1$ layers of neurons with $L \geq 1$. Each layer $l \in \{0, 1, ..., L\}$ is composed of $n_l$ neurons, numbered from $1$ to $n_l$. These layers and neurons define the architecture of the NN. Layer $0$ is fictitious and corresponds to the input of the NN, while the last layer $L$ corresponds to its outputs. Layers $1$ to $L-1$ are typically referred to as hidden layers. Let $x^l_i$ be the output of the $i$th neuron of the $l$th layer, with $i \in \{1,...,n_l\}$. The inputs to the NN can be represented as $x^0_i$ or simply $x_i$. Moreover, we represent the outputs as $x^L_i$ or simply $o_i$.

The values $x^l_i$ of the neurons in a given layer $l$ are computed through the output values $x^{l-1}_j$ of the previous layer, with $j \in \{1,...,n_{l-1}\}$. Each neuron applies a linear combination of the output of the neurons in the previous layer. Then, the neuron applies a nonlinear function, also known as an activation function. The output of the linear part is represented as $\sum_{j=1}^{n_{l-1}} w^{l}_{i,j} x^{l-1}_{j} + b^{l}_{i}$ where $w^{l}_{i,j}$ and $b^{l}_{i}$ denote the weights and bias, respectively, serving as parameters of the $i$th neuron of layer $l$. In this work, we consider only the Rectified Linear Unit ($\mathrm{ReLU}$) as activation function because it can be represented by linear constraints due to its piecewise-linear nature. This function is a widely used activation whose output is the maximum between its input value and zero. Then, $x^{l}_{i} = \mathrm{ReLU}(\sum_{j=1}^{n_{l-1}} w^{l}_{i,j} x^{l-1}_{j} + b^{l}_{i})$ is the output of the $\mathrm{ReLU}$. The last layer $L$ is composed of $n_L = k$ neurons, one for each class. It is common to normalize the output layer using an additional Softmax layer. Consequently, these values represent the probabilities associated with each class. The class with the highest probability is chosen as the predicted class. However, we do not need to consider this normalization transformation as it does not change the maximum value of the last layer. Thus, the predicted class is $c_i \in \mathcal{K}$ such that $i = \argmax_{j \in \{1, ..., k\}} x^L_j$.

\subsection{MILP – Mixed Integer Linear Programming}

In Mixed Integer Linear Programming (MILP), the objective is to optimize a linear function subject to linear constraints, where some or all of the variables are required to be integers \cite{milp1971}. To illustrate the structure of a MILP problem, we provide an example below:
\footnotesize
\begin{align}
\label{eq:milp}
%\begin{aligned}
\min \quad & y_1 \notag\\
\textrm{s.t.} \quad & 1 \leq x_1 \leq 3 \notag\\
                    %&0 \leq x_2 \leq 1\\
                    %& 3x_1 + x_2 + s_1 - 2 = y_1 \\
                    & 3x_1 + s_1 - 2 = y_1 \notag\\
                    %& 0 \leq y_1 \leq 3x_1 + x_2 - 2 \\
                    & 0 \leq y_1 \leq 3x_1 - 2 \notag\\
                    & 0 \leq s_1 \leq 3x_1 - 2\\
                    & z_1 = 1 \to y_1 \leq 0 \notag\\
                    & z_1 = 0 \to s_1 \leq 0 \notag\\
                    & z_1 \in \{0, 1\} \notag
%\label{eq:milp_example4} 
%\end{aligned}
\end{align}\normalsize
% \begin{equation}
% \label{eq:milp}
% \begin{aligned}
% \min \quad & y_1 \\
% \textrm{s.t.} \quad & 1 \leq x_1 \leq 3\\
%                     %&0 \leq x_2 \leq 1\\
%                     %& 3x_1 + x_2 + s_1 - 2 = y_1 \\
%                     & 3x_1 - 2 \leq y_1 \\
%                     %& 0 \leq y_1 \leq 3x_1 + x_2 - 2 \\
%                     & y_1 \leq 3x_1 - 2 - 0.5(1-z_1)\\
%                     & 0 \leq y_1 \leq 8 z_1 \\
%                     %& 0 \leq s_1 \leq 3x_1 + x_2 - 2\\
%                     %& 0 \leq s_1 \leq 3x_1 - 2\\
%                     %& z_1 = 1 \to y_1 \leq 0 \\
%                     %& z_1 = 0 \to s_1 \leq 0 \\
%                     & z_1 \in \{0, 1\}
% %\label{eq:milp_example4} 
% \end{aligned}
% \end{equation}

%MILP is a crucial technique in our work for determining the lower and upper bounds of each neuron in the NNs. For example, we utilize a minimization problem to determine the lower bound of neurons within NNs. This process involves formulating an objective function that seeks to minimize the lower bound, subject to constraints that reflect the behaviour of NNs. 

In the MILP problem in (\ref{eq:milp}), we want to ﬁnd values for variables $x_1, y_1, s_1, z_1$ minimizing the value of the objective function $y_1$, among all values that satisfy the constraints. Variable $z_1$ is binary since $z_1 \in \{0, 1\}$ is a constraint in the MILP, while variables $x_1, y_1, s_1$ have the real numbers $\mathbb{R}$ as their domain. The constraints in a MILP may appear as linear equations, linear inequalities, and indicator constraints. Indicator constraints can be seen as logical implications of the form $z = v \to \sum^n_{i=1} w_i x_i \leq b$ such that $z$ is a binary variable, and $v$ is a constant $0$ or $1$.%\cite{bonami2015mathematical}.

An important observation is that a MILP problem without an objective function corresponds to a satisfiability problem, as discussed in Subsection~\ref{subsec:logic}. Given that the approach for computing explanations relies on logical consequence, and considering the connection between satisfiability and logical consequence, we employ a MILP solver to address explanation tasks. Additionally, throughout the construction of the MILP model, we may utilize optimization, specifically employing a MILP solver, to determine tight lower and upper bounds for the neurons of NNs.

% In this project, we utilize satisfaction models to find explications and optimization models to determine the most precise bounds. It is crucial to emphasize that the objective function is essential in optimization models, whereas it is not required in satisfaction models. Additionally, we also use external solvers as oracles to solve the MILP.

\section{Logic-based XAI for Neural Networks}

%As previously mentioned, most methods for explaining ML models are heuristic, resulting in explanations that can not be fully trusted. Logic-based explainability provides results with guarantees of correctness and irredundancy. For NNs, Ignatiev, Narodytska, and Marques-Silva \cite{ignatiev2019abduction} employed a logic-based encoding based on linear constraints with binary and continuous variables. 

As previously mentioned, most methods for explaining ML models are heuristic, resulting in explanations that can not be fully trusted. Logic-based explainability provides results with guarantees of correctness and irredundancy. For NNs, \citet{ignatiev2019abduction} employed a logic-based approach based on linear constraints with binary and continuous variables. Given an instance, this approach identifies a subset of input features sufficient to justify the correspondent output given by the NN. This method ensures correctness and minimality of explanations, which are referred to as \textit{abductive explanations}. An abductive explanation is a subset of features that form a rule. When applied, this rule guarantees the same model prediction. The formal definition below encapsulates this notion.

\begin{definition}[Abductive Explanation \cite{ignatiev2019abduction}]
Let $\mathcal{I} = \{x_1 = v_1, ..., x_n = v_n\}$ be an instance and $\mathcal{N}$ be a NN such that $\mathcal{N}(\mathcal{I}) = c \in \mathcal{K}$. An \emph{abductive explanation} $\mathcal{X}$ is a minimal subset of $\mathcal{I}$ such that for all $v_1' \in [l_1, u_1], ..., v_n' \in [l_n, u_n]$, if $v'_j = v_j$ for each $x_j = v_j \in \mathcal{X}$, then $\mathcal{N}(\{x_1 = v'_1, ..., x_n = v'_n\}) = c$.
\end{definition}

An abductive explanation is a minimal set of features from an instance that are sufficient for the prediction. This subset ensures that, when the values of these features are fixed and the other features are varied within their possible ranges, the prediction remains the same. In other words, it identifies key features that are sufficient for the output. Minimality ensures that the explanation $\mathcal{X}$ does not include any redundant features. In other words, removing any feature $x_j = v_j$ from $\mathcal{X}$ would result in a subset that no longer guarantees the same prediction $c$ over the defined bounds.

It is important to note that a given instance may have multiple distinct abductive explanations. Different subsets of features may independently satisfy the conditions for an explanation while maintaining minimality. This means that there can be multiple ways to justify a prediction, each highlighting different aspects of the instance that are sufficient to ensure the same output.

The approach by \citet{ignatiev2019abduction} to obtain abductive explanations for NNs works as follows. First, the NN and the input domain $\mathcal{D}$ are encoded as a set of formulas $\mathcal{F}$. We encode a NN with $L+1$ layers and $\mathcal{D}$ as $\mathcal{F}$ in (\ref{eq:input})-(\ref{eq:indicator2}), where $l= 1,\dots,L-1$ and, for each $l$, $j = 1,\dots,n_l$.\footnotesize
\begin{align}
     &lb_i \leq x_{i} \leq ub_i,\quad i = 1, ..., n \label{eq:input}\\
     &\sum_{i=1}^{n_{l-1}} w^{l}_{i,j} x^{l-1}_i + b^{l}_{j} = x^{l}_{j} - s^{l}_{j} \label{eq:indicator1}\\
         &z^{l}_{j} = 1 \rightarrow x^{l}_{j} \leq 0 \\
         &z^{l}_{j} = 0 \rightarrow s^{l}_{j} \leq 0 \\
         &z^{l}_{j}=0 \vee z^{l}_{j}=1 \label{eq:z_or} \\
         &0 \leq x^{l}_{j} \leq ub^{l}_{x,j} \label{eq:ub} \\
         &0 \leq s^{l}_{j} \leq ub^{l}_{s,j} \label{eq:ubs}\\
         &o_i = \sum_{j=1}^{n_{L-1}} w^{L}_{i,j} x^{L-1}_i + b^{L}_{i},\quad i = 1, ..., k \label{eq:indicator2}
\end{align}\normalsize

In the following, we explain the notation. The encoding uses variables $x^{l}_{j}$ and $o_i$ with the same meaning as in the notation for NNs. Auxiliary variables $s^{l}_{j}$ and $z^{l}_{j}$ control the behaviour of $\mathrm{ReLU}$ activations. Moreover, this encoding uses implications to represent the behavior of $\mathrm{ReLU}$. Variable $z^{l}_{j}$ is binary and if $z^{l}_{j}$ is equal to $1$, the $\mathrm{ReLU}$ output $x^{l}_{j}$ is $0$ and $- s^{l}_{j}$ is equal to the linear part. Otherwise, the output $x^{l}_{j}$ is equal to the linear part and $s^{l}_{j}$ is equal to $0$. The bounds $ub^{l}_{x,j}$ are defined by isolating variable $x^{l}_{j}$ from other constraints in subsequent layers. Then, $x^{l}_{j}$ is maximized to find its upper bound. A similar process is applied to find the bounds $ub^{l}_{s,j}$ for variables $s^{l}_{j}$. In this case, $ub^{l}_{s,j}$ corresponds to the absolute value of the smallest non-positive input to the ReLU function. These bounds $ub^{l}_{x,j}$ and $ub^{l}_{s,j}$ can be determined due to the bounds of the features. Furthermore, these bounds can assist the solver in accelerating the computation of explanations. Finally, in (\ref{eq:input}), variables $x_{i}$ have lower and upper bounds $lb_i$, $ub_i$, respectively, defined by the input domain. Therefore, $\mathcal{D} \subseteq \mathcal{F}$.
%The constant $ub^{l}_{s,i}$ is the upper bound of variable $s^{l}_{i}$, and the constant $ub^{l}_{x,i}$ is the upper bound of variable $x^{l}_{i}$.

The second step in this approach consists of defining a formula $E_j$ that captures the prediction of the NN. Given an instance $\mathcal{I} = \{x_1 = v_1, x_2 = v_2, ..., x_n = v_n\}$ from the input domain, the neural network $\mathcal{N}$, and its corresponding prediction $\mathcal{N}(\mathcal{I}) = c_j$, we encode this information by asserting that $o_j$ is the largest among all output neurons: $E_j = \bigwedge_{i=1, i \neq j}^{k} o_j > o_i$. Then, $\mathcal{I} \cup \mathcal{F}$ is satisfiable and $\mathcal{I} \cup \mathcal{F} \models E_j$. An abductive explanation $\mathcal{X}$ is calculated removing feature by feature from $\mathcal{I}$. For example, given $x_i=v_i \in \mathcal{I}$, if $(\mathcal{I} \setminus \{x_i=v_i\}) \cup \mathcal{F} \models E_j$, then feature $x_i = v_i$ is not required for the explanation and can be removed. In other words, $x_i$ can take any value within its domain, while guaranteeing that the prediction remains $c_j$. Otherwise, if $(\mathcal{I} \setminus \{x_i=v_i\}) \cup \mathcal{F} \not\models E_j$, then $x_i = v_i$ is kept, since class $c_j$ cannot be guaranteed. In this case, there exists some value within the domain of $x_i$ that would change the prediction. This process is described in Algorithm \ref{algorithm1} and is performed for all features. Then, $\mathcal{X}$ is the result at the end of this procedure. This means that for the values of the features in $\mathcal{X}$, the NN makes the same classification $c_j$, whatever the values of the remaining features.

\begin{algorithm}
\footnotesize
\caption{Computing an Explanation X} \label{algorithm1}
\begin{algorithmic}[1]
\Require{$\mathcal{F}$, $\mathcal{I}$, $E_j$}
%\Ensure{minimal explanation $C_m$}
\State $\mathcal{X} \gets \mathcal{I}$\;
\For{$x_i=v_i$ \text{ in } $\mathcal{I}$}
\If{$(\mathcal{X} \setminus \{x_i=v_i\}) \cup F \models E_j$}
\State $\mathcal{X} \gets \mathcal{X} \backslash \{x_i=v_i\}$\
\EndIf
\EndFor
\State \Return $\mathcal{X}$
%   %}
\end{algorithmic}
\normalsize
\end{algorithm}

As stated in (\ref{ent_unsat}), to check entailments $(\mathcal{X} \setminus \{x_i=v_i\}) \cup \mathcal{F} \models E_j$, it is equivalent to test whether $(\mathcal{X} \setminus \{x_i=v_i\}) \cup \mathcal{F} \cup \{\neg E_j\}$ is unsatisfiable. Since $\mathcal{F}$, $\mathcal{X}$ and $\neg E$ are encoded as linear constraints with continuous and binary variables, along with indicator constraints, such entailments can be addressed using a MILP solver.

\section{Logic-based XAI for NNs via Slicing}

Previous work \cite{ignatiev2019abduction} introduced a logic-based approach for explaining NNs, identifying a subset of input features sufficient to justify a given output. This method ensures correctness and minimality of explanations, which are referred to as abductive explanations. Unfortunately, this approach becomes computationally prohibitive with increasing network size. In this work, we address this limitation by exploring the concept of \emph{slicing}. We partition the feature domains into smaller subdomains. Then, within each subdomain, we can simplify the original NN, potentially reducing the time to compute explanations. Therefore, domain slicing may enable the computation of abductive explanations for larger NNs.

There are several approaches for partitioning the input domain. One direct method evaluated in our work involves dividing the domain of a feature into two equal sub-domains. More formally, let $x_i = v_i$ in $\mathcal{I}$ such that $lb_i \leq x_i \leq ub_i$ in $F$ and define $m_i := (l_i + u_i)/2$. Then, we consider the sub-domains $lb_i \leq x_i \leq m_i$ and $m_i \leq x_i \leq ub_i$ for feature $x_i$. To illustrate the method, we consider Example~\ref{ex:slicing_concrete} below.

\begin{figure}[t]%htbp
\centering
\begin{tikzpicture}[x=2.5cm, y=1.5cm, >=stealth,scale=0.8]
% Camada de entrada
\node[circle, draw, label=above:{$[0.2, 0.7]$}] (I1) at (0,1) {$x_1$};
\node[circle, draw, label=below:{$[0.2, 0.5]$}] (I2) at (0,-1) {$x_2$};

% Camada oculta antes da ReLU
\node[circle, draw, label=above:{$[-0.3, 0.8]$}] (H1_pre) at (1,1) {$y_1$};
\node[circle, draw, label=below:{$[-0.3, 0.5]$}] (H2_pre) at (1,-1) {$y_2$};

% Camada oculta depois da ReLU
\node[circle, draw] (H1_post) at (2,1) {$x_1^1$};
\node[circle, draw] (H2_post) at (2,-1) {$x_2^1$};

% Camada de saída
\node[circle, draw] (O1) at (3,1) {$o_1$};
\node[circle, draw] (O2) at (3,-1) {$o_2$};

% Conexões camada de entrada -> camada oculta antes da ReLU
%[pos=0.25, above]
\draw[->] (I1) -- (H1_pre) node[midway, above] {$-1$};
\draw[->] (I1) -- (H2_pre) node[pos=0.25, above] {$1$};
\draw[->] (I2) -- (H1_pre) node[pos=0.25, below] {$2$};
\draw[->] (I2) -- (H2_pre) node[midway, above] {$-1$};

% Conexões camada oculta antes da ReLU -> depois da ReLU
\draw[->] (H1_pre) -- (H1_post) node[midway, above] {\text{ReLU}};
\draw[->] (H2_pre) -- (H2_post) node[midway, above] {\text{ReLU}};

% Conexões camada oculta depois da ReLU -> camada de saída
\draw[->] (H1_post) -- (O1) node[midway, above] {$1$};
\draw[->] (H1_post) -- (O2) node[pos=0.25, above] {$1$};
\draw[->] (H2_post) -- (O1) node[pos=0.25, below] {$1$};
\draw[->] (H2_post) -- (O2) node[midway, above] {$-1$};

\end{tikzpicture}
\caption{Example of Neural Network}
\label{fig1}
\end{figure}
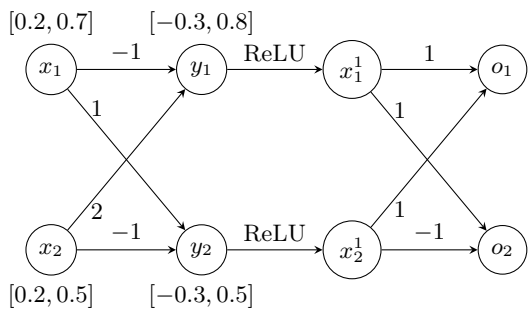

\begin{example}\label{ex:slicing_concrete}
For instance, consider an input domain with two features:  
$\mathcal{D} = \{0.2 \leq x_1 \leq 0.7, 0.2 \leq x_2 \leq 0.5\}.$ Also consider the simple NN in Fig.~\ref{fig1}. This NN consists of an input layer, one hidden layer, and one output layer. Each layer consists of two neurons. Then, there are two inputs to this NN, $x_1$ with domain $[0.2, 0.7]$ and $x_2$ with domain $[0.2, 0.5]$. To simplify the example, we separate each neuron in the hidden layer into two parts: one representing the output of the linear transformation and the other capturing the output of the ReLU activation. We denote the upper neuron in the hidden layer as $y_1$ and the lower neuron as $y_2$. The weights for the linear transformations are represented by weights on the edges. The bias for each neuron is $0$ and has been omitted to simplify the example. Above or below some neurons, there is an interval representing a tight range of values the neuron can take. These initial intervals for $y_1$ and $y_2$ can be obtained through optimization, for example. In particular, the upper bound of $y_1$ can be determined by maximizing: $-1 x_1 + 2 x_2.$ A lower bound for $y_1$ can be found using a similar optimization process. This optimization is feasible due to the predefined bounds on the input features. The upper bound $ub^{1}_{x,1}$ of $x_1^1$  is, then, the maximum between $0$ and the largest input to the ReLU function. In our example, this results in $ub^{1}_{x,1} = max\{0, 0.8\} = 0.8.$ Similarly, the bound $ub^{1}_{s,1}$ of $s_1^1$ is the absolute value of the minimum between $0$ and the smallest input to the ReLU function. In this case, we obtain: $ub^{1}_{s,1} = \left| \min\{0, -0.3\} \right| = 0.3.$ A similar reasoning applies to $x_2^1$ and $s_2^1$, where their respective bounds can be determined using the same approach based on the inputs to the ReLU function.

In this example, we apply the slicing method to \(x_2\), obtaining the two subdomains: $\mathcal{D}_1 = \{0.2 \leq x_1 \leq 0.7, 0.2 \leq x_2 \leq 0.35\}$ and $\mathcal{D}_2 = \{0.2 \leq x_1 \leq 0.7, 0.35 \leq x_2 \leq 0.5\}$. Then, within each subdomain, we can simplify the original NN, potentially reducing the time to compute explanations. For example, the simplification corresponding to subdomain $\mathcal{D}_1$ is presented in Fig.~\ref{fig:nn_first_simpl}. 

The subdomain $\mathcal{D}_1$ affects the propagation of values within the NN. For $y_1$, its original interval was $[-0.3,0.8]$. With the new subdomain restriction, the values of $x_2$ are now constrained within the reduced range $[0.2,0.35]$. This decreases the influence of $x_2$ on the linear transformations, lowering the upper bound of $y_1$ to $0.5$. Thus, the updated interval for $y_1$ becomes $[-0.3,0.5]$. Consequently, $ub^{1}_{x,1} = max\{0, 0.5\} = 0.5$. Therefore, for subdomain $\mathcal{D}_1$, the formula $0 \leq x^{1}_{1} \leq 0.8$ corresponding to Equation~(\ref{eq:ub}) for $l=1$ and $j=1$, can be simplified to $0 \leq x^{1}_{1} \leq 0.5.$ In a MILP problem, tightening the bounds of a continuous variable reduces the feasible region, thereby limiting the search space explored by the solver and potentially improving solver efficiency.

For $y_2$, whose original interval was $[-0.3,0.5]$, the new restriction on $x_2$ in subdomain $\mathcal{D}_1$ affects the lower bound. Since the minimum value of $x_2$ is now $0.2$, the linear transformation results in an updated lower bound of $-0.15$ for $y_2$. Consequently, the new interval for $y_2$ becomes $[-0.15,0.5]$. Therefore, $ub^{1}_{s,2} = \left| \min\{0, -0.15\} \right| = 0.15$. Then, for subdomain $\mathcal{D}_1$, the formula $0 \leq s^{1}_{2} \leq 0.3$ corresponding to Equation~(\ref{eq:ubs}) for $l=1$ and $j=1$, can be simplified to $0 \leq s^{1}_{2} \leq 0.15.$

\begin{figure}[tbp]%htbp
\centering
\begin{tikzpicture}[x=2.5cm, y=1.5cm, >=stealth,scale=0.8]
% Camada de entrada
\node[circle, draw, label=above:{$[0.2, 0.7]$}] (I1) at (0,1) {$x_1$};
\node[circle, draw, label=below:{$[0.2, 0.35]$}] (I2) at (0,-1) {$x_2$};

% Camada oculta antes da ReLU
\node[circle, draw, label=above:{$[-0.3, 0.5]$}] (H1_pre) at (1,1) {$y_1$};
\node[circle, draw, label=below:{$[-0.15, 0.5]$}] (H2_pre) at (1,-1) {$y_2$};

% Camada oculta depois da ReLU
\node[circle, draw] (H1_post) at (2,1) {$x_1^1$};
\node[circle, draw] (H2_post) at (2,-1) {$x_2^1$};

% Camada de saída
\node[circle, draw] (O1) at (3,1) {$o_1$};
\node[circle, draw] (O2) at (3,-1) {$o_2$};

% Conexões camada de entrada -> camada oculta antes da ReLU
%[pos=0.25, above]
\draw[->] (I1) -- (H1_pre) node[midway, above] {$-1$};
\draw[->] (I1) -- (H2_pre) node[pos=0.25, above] {$1$};
\draw[->] (I2) -- (H1_pre) node[pos=0.25, below] {$2$};
\draw[->] (I2) -- (H2_pre) node[midway, above] {$-1$};

% Conexões camada oculta antes da ReLU -> depois da ReLU
\draw[->] (H1_pre) -- (H1_post) node[midway, above] {\text{ReLU}};
\draw[->] (H2_pre) -- (H2_post) node[midway, above] {\text{ReLU}};

% Conexões camada oculta depois da ReLU -> camada de saída
\draw[->] (H1_post) -- (O1) node[midway, above] {$1$};
\draw[->] (H1_post) -- (O2) node[pos=0.25, above] {$1$};
\draw[->] (H2_post) -- (O1) node[pos=0.25, below] {$1$};
\draw[->] (H2_post) -- (O2) node[midway, above] {$-1$};

\end{tikzpicture}
\caption{Slicing respective to subdomain $\mathcal{D}_1$}
\label{fig:nn_first_simpl}
\end{figure}
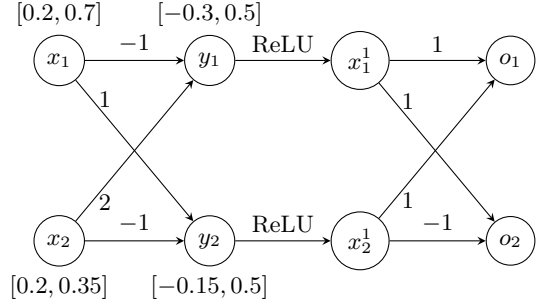

%\textcolor{red}{comentar sobre o $-0.3$ e o 0.3, que o s$^1_1$ é o maior vlaor negativo que pode entrar na RELU.}

\begin{figure}[tbp]%htbp
\centering
\begin{tikzpicture}[x=2.5cm, y=1.5cm, >=stealth,scale=0.8]
% Camada de entrada
\node[circle, draw, label=above:{$[0.2, 0.7]$}] (I1) at (0,1) {$x_1$};
\node[circle, draw, label=below:{$[0.35, 0.5]$}] (I2) at (0,-1) {$x_2$};

% Camada oculta antes da ReLU
\node[circle, draw, label=above:{$[0.0, 0.8]$}] (H1_pre) at (1,1) {$y_1$};
\node[circle, draw, label=below:{$[-0.3, 0.35]$}] (H2_pre) at (1,-1) {$y_2$};

% Camada oculta depois da ReLU
\node[circle, draw] (H1_post) at (2,1) {$x_1^1$};
\node[circle, draw] (H2_post) at (2,-1) {$x_2^1$};

% Camada de saída
\node[circle, draw] (O1) at (3,1) {$o_1$};
\node[circle, draw] (O2) at (3,-1) {$o_2$};

% Conexões camada de entrada -> camada oculta antes da ReLU
%[pos=0.25, above]
\draw[->] (I1) -- (H1_pre) node[midway, above] {$-1$};
\draw[->] (I1) -- (H2_pre) node[pos=0.25, above] {$1$};
\draw[->] (I2) -- (H1_pre) node[pos=0.25, below] {$2$};
\draw[->] (I2) -- (H2_pre) node[midway, above] {$-1$};

% Conexões camada oculta antes da ReLU -> depois da ReLU
\draw[->] (H1_pre) -- (H1_post) node[midway, above] {\text{ReLU}};
\draw[->] (H2_pre) -- (H2_post) node[midway, above] {\text{ReLU}};

% Conexões camada oculta depois da ReLU -> camada de saída
\draw[->] (H1_post) -- (O1) node[midway, above] {$1$};
\draw[->] (H1_post) -- (O2) node[pos=0.25, above] {$1$};
\draw[->] (H2_post) -- (O1) node[pos=0.25, below] {$1$};
\draw[->] (H2_post) -- (O2) node[midway, above] {$-1$};

\end{tikzpicture}
\caption{Slicing respective to subdomain $\mathcal{D}_2$}
\label{fig:nn_sec_simpl}
\end{figure}
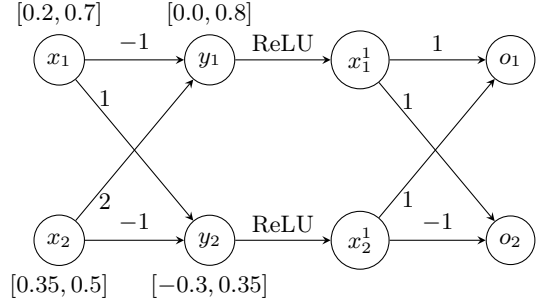

The simplification corresponding to subdomain $\mathcal{D}_2$ is presented in Fig.~\ref{fig:nn_sec_simpl}. Initially, the interval of neuron $y_2$ was $[-0.3,0.5]$. After slicing, due to the new range of $x_2$ in the subdomain $\mathcal{D}_2$, the upper bound of $y_2$ decreases to $0.35$. The lower bound of $y_2$ remains unchanged at $-0.3$. 

The most interesting case in this example occurs with neuron $y_1$ in subdomain $\mathcal{D}_2$. Initially, its interval was $[-0.3,0.8]$, but with the subdomain constraint $x_2 \geq 0.35$ in $\mathcal{D}_2$, the lower bound of $y_1$ increases from $-0.3$ to $0.0$. In this case, the formulas $-1 x_1 + 2 x_2 = x^{1}_{1} - s^{1}_{1}$, $z^{1}_{1} = 1 \rightarrow x^{1}_{1} \leq 0$, $z^{1}_{1} = 0 \rightarrow s^{1}_{1} \leq 0$, $z^{1}_{1}=0 \vee z^{1}_{1}=1$, $0 \leq x^{l}_{j} \leq 0.8$, and $0 \leq s^{l}_{j} \leq 0.3$ corresponding to Equations~(\ref{eq:indicator1})-(\ref{eq:z_or}) for $l=1$ and $j=1$, i.e., for neuron $x_1^1$, can be simplified to $-1 x_1 + 2 x_2 = x^{1}_{1}$ and $0 \leq x^{1}_{1} \leq 0.8$. In other words, we remove the constraints associated with binary variable $z_1^1$ and real variable $s_1^1$. Since MILP solvers are generally more efficient with fewer binary variables, this reduction potentially improves computational efficiency.
%\begin{align*}
%     -&1 x_1 + 2 x_2 = x^{1}_{1} \\
%     &0 \leq x^{1}_{1} \leq 0.8
%\end{align*}\normalsize

%This results in two NNs, one for each subdomain.
%\begin{example}
%To illustrate the method, we 
%If the network had additional layers, the values of $x_7$ and $x_8$ would pass through the ReLU activation function, and their resulting values would propagate to the subsequent layers.    
%\end{example}
\end{example}

%\textcolor{red}{Falar que vamos continuar explorando a ideia de subdomains agora fazendo slicing em duas features}.

We will now continue exploring the slicing approach by extending it to two features simultaneously. This will allow us to understand how the subdomains are formed and how the combined effect of slicing multiple features partitions the input domain.

\begin{example}\label{ex:slicing1}
Consider an input domain with three features: $\mathcal{D} = \{lb_1 \leq x_1 \leq ub_1, lb_2 \leq x_2 \leq ub_2, lb_3 \leq x_3 \leq ub_3\}.$ Applying this slicing method to \(x_1\), we obtain two sub-domains: $\mathcal{D}_1 = \{lb_1 \leq x_1 \leq m_1, lb_2 \leq x_2 \leq ub_2, lb_3 \leq x_3 \leq ub_3\}$ and $\mathcal{D}_2 = \{m_1 \leq x_1 \leq ub_1, lb_2 \leq x_2 \leq ub_2, lb_3 \leq x_3 \leq ub_3\}.$ Now, applying the same partitioning method to \(x_2\), each of the previous sub-domains is further divided into two, resulting in four sub-domains: $\mathcal{D}_1 = \{lb_1 \leq x_1 \leq m_1, lb_2 \leq x_2 \leq m_2, lb_3 \leq x_3 \leq ub_3\}$, $\mathcal{D}_2 = \{lb_1 \leq x_1 \leq m_1, m_2 \leq x_2 \leq ub_2, lb_3 \leq x_3 \leq ub_3\}$, $\mathcal{D}_3 = \{m_1 \leq x_1 \leq ub_1, lb_2 \leq x_2 \leq m_2, lb_3 \leq x_3 \leq ub_3\}$, and $\mathcal{D}_4 = \{m_1 \leq x_1 \leq ub_1, m_2 \leq x_2 \leq ub_2, lb_3 \leq x_3 \leq ub_3\}$. By continuing this process, we iteratively refine the input domain, generating a finer partitioning of the domain.
\end{example}

In networks with a larger number of input neurons, the number of resulting input subdomains can become prohibitively large. Then, we will evaluate the domain slicing technique by limiting the values to a maximum of three features.

%\textcolor{red}{AGORA SAO AS DEMONSTRACOES}
%, already simplified
%partitioning of the input space.
Next, we explore how analyzing a single subdomain can sometimes be enough to determine that a feature must be retained in the explanation. This occurs when the constraints within that subdomain inherently enforce the necessity of the feature to ensure the prediction, even without considering the entire input domain.

In Algorithm~\ref{algorithm1}, we test whether $(\mathcal{X} \setminus {x_i=v_i}) \cup \mathcal{F} \models E_j$. Since the domain of feature $x_i$ is in $\mathcal{F}$, i.e., $l_i \leq x_i \leq u_i \in \mathcal{F}$, we are effectively testing whether all assignments $\mathcal{A}$ that satisfy $ (\mathcal{X} \setminus {x_i=v_i})$, $\mathcal{F}$ and $l_i \leq x_i \leq u_i$ also satisfy $E_j$. If there exists an assignment $\mathcal{A}$ where $x_i \mapsto v_i' \in \mathcal{A}$ for some $v_i' \in [l_i, u_i]$ such that $\mathcal{A}$ satisfy $(\mathcal{X} \setminus {x_i=v_i})$ and $\mathcal{F}$ but does not satisfy $E_j$, then 
$(\mathcal{X} \setminus {x_i=v_i}) \cup \mathcal{F} \not\models E_j.$ Consequently, Algorithm~\ref{algorithm1} ensures that $x_i = v_i$ remains in $\mathcal{X}$. 

Now, suppose we apply slicing to feature $x_i$, splitting its domain into two subdomains $\mathcal{D}_1$ and $\mathcal{D}_2$ where $l_i \leq x_i \leq m_i \in \mathcal{D}_1$ and $m_i \leq x_i \leq u_i \in \mathcal{D}_2$. To verify whether $(\mathcal{X} \setminus {x_i=v_i}) \cup \mathcal{F} \models E_j$ it suffices to check both subdomains separately: $(\mathcal{X} \setminus {x_i=v_i}) \cup \mathcal{F} \cup \{m_i \leq x_i\} \models E_j$ and $(\mathcal{X} \setminus {x_i=v_i}) \cup \mathcal{F} \cup \{x_i \leq m_i\} \models E_j$. Since $l_i \leq x_i$ and $x_i \leq u_i$ are already present in $\mathcal{F}$, if 
$(\mathcal{X} \setminus {x_i=v_i}) \cup F \cup \{x_i \leq m_i\} \not\models E_j,$ then there exists an assignment $\mathcal{A}$ with $x_i \mapsto v'_i \in \mathcal{A}$ for some $v'_i \in [l_i, m_i]$ such that $\mathcal{A}$ satisfy $(\mathcal{X} \setminus {x_i=v_i})$ and $\mathcal{F}$ but does not satisfy $E_j$. Consequently, since $v'_i \in [l_i, m_i] \subseteq [l_i, u_i]$, the same assignment $\mathcal{A}$ satisfies $(\mathcal{X} \setminus {x_i=v_i})$ and $\mathcal{F}$ but does not satisfy $E_j$. Thus, we conclude that $(\mathcal{X} \setminus {x_i=v_i}) \cup \mathcal{F} \not\models E_j$ without needing to evaluate the entire domain $l_i \leq x_i \leq u_i$.

Similarly, if $(\mathcal{X} \setminus \{x_i=v_i\}) \cup \mathcal{F} \cup \{m_i \leq x_i\} \not\models E_j$, the same reasoning applies, ensuring that $(\mathcal{X} \setminus {x_i=v_i}) \cup \mathcal{F} \not\models E_j$. We now consider the converse. Assume that $(\mathcal{X} \setminus {x_i=v_i}) \cup \mathcal{F} \not\models E_j$. Then there exists an assignment $\mathcal{A}$ where $x_i \mapsto v'_i \in \mathcal{A}$ for some $v'_i \in [l_i, u_i]$, such that $\mathcal{A}$ satisfies $(\mathcal{X} \setminus \{x_i = v_i\})$ and $\mathcal{F}$, but does not satisfy $E_j$. We have two possible cases (both may hold if $v'_i = m_i$). If $v'_i \in [l_i, m_i]$, then it follows that $(\mathcal{X} \setminus \{x_i = v_i\}) \cup \mathcal{F} \cup \{x_i \leq m_i\} \not\models E_j$. If $v'_i \in [m_i, u_i]$, then we conclude that $(\mathcal{X} \setminus \{x_i = v_i\}) \cup \mathcal{F} \cup \{m_i \leq x_i\} \models E_j$.

From the reasoning above, we conclude that verifying $(\mathcal{X} \setminus \{x_i = v_i\}) \cup \mathcal{F} \models E_j$ can be reduced to checking the two subdomains separately. Furthermore, we established that if $(\mathcal{X} \setminus \{x_i = v_i\}) \cup \mathcal{F} \not\models E_j$, then at least one of these subdomain checks must also fail. These observations lead to the following proposition:

\begin{proposition}\label{proposition:slicing}
Let $m_i := (l_i + u_i)/2$ for a feature $x_i$ with $l_i \leq x_i \leq u_i \in \mathcal{D}$. Then, $(\mathcal{X} \setminus \{x_i = v_i\}) \cup \mathcal{F} \models E_j$ if and only if $(\mathcal{X} \setminus \{x_i = v_i\}) \cup \mathcal{F} \cup \{x_i \leq m_i\} \models E_j$ and $(\mathcal{X} \setminus \{x_i = v_i\}) \cup \mathcal{F} \cup \{m_i \leq x_i\} \models E_j.$
\end{proposition}

\noindent This observation naturally extends to multiple features. Specifically, if $(\mathcal{X} \setminus \{x_i=v_i, x_p=v_p\}) \cup \mathcal{F} \cup \{ x_i \leq m_i, m_p \leq x_p\} \not\models E_j,$ then considering the entire domain also does not entail $E_j$, i.e., $(\mathcal{X} \setminus \{x_i=v_i, x_p=v_p\}) \cup \mathcal{F} \not\models E_j.$ This suggests a structured way to refine the input domain. The following proposition formalizes this idea for two features, showing that verifying entailment over the full domain can be reduced to verifying it over smaller subdomains.

\begin{proposition}\label{proposition:slicing2}
Let $m_i := (l_i + u_i)/2$ for a feature $x_i$ with $l_i \leq x_i \leq u_i \in \mathcal{D}$, and $m_p := (l_p + u_p)/2$ for a feature $x_p$ with $l_p \leq x_p \leq u_p \in \mathcal{D}$. Then, $(\mathcal{X} \setminus \{x_i = v_i, x_p = v_p\}) \cup \mathcal{F} \models E_j$ if and only if $(\mathcal{X} \setminus \{x_i = v_i, x_p = v_p\}) \cup \mathcal{F} \cup \{x_i \leq m_i, x_p \leq m_p\} \models E_j$, $(\mathcal{X} \setminus \{x_i = v_i, x_p = v_p\}) \cup \mathcal{F} \cup \{x_i \leq m_i, m_p \leq x_p\} \models E_j$, $(\mathcal{X} \setminus \{x_i = v_i, x_p = v_p\}) \cup \mathcal{F} \cup \{m_i \leq x_i, x_p \leq m_p\} \models E_j$, and $(\mathcal{X} \setminus \{x_i = v_i, x_p = v_p\}) \cup \mathcal{F} \cup \{m_i \leq x_i, m_p \leq x_p\} \models E_j$.
\end{proposition}

\noindent The proposition shows that instead of evaluating entailment over the entire input space, it suffices to check it over a partitioned set of subdomains. Moreover, each subdomain can simplify the constraints of the MILP problem, as demonstrated in Example~\ref{ex:slicing_concrete}. However, in NNs with a large number of input features, the number of resulting subdomains can grow exponentially, making the approach computationally impractical. To address this, in our experiments, we limit the domain slicing technique to at most three features, which are selected incrementally and at random. In the subsequent section, we will evaluate the impact of domain slicing by measuring its effect on explanation computation time.

\section{Experiments}
%To evaluate the performance of each slicing variation, we compared the total explanation time using a fixed set of randomly selected instances. The total time is obtained by summing the explanation times for all instances. 

%Additionally, for each slicing level, we counted the number of binary variables removed due to the input domain and subdomains. Specifically, for each subdomain, we counted the number of binary variables removed and computed the average across all subdomains.

In our experiments, we applied slicing to continuous input features. For each trained NN, we considered four slicing cases: no slicing and one to three sliced attributes. The case without any slicing represents the baseline model found in other works \cite{ignatiev2019abduction}, whereas the cases with one or more slicings represent the approach we propose. The features selected for slicing were chosen randomly, and each additional slice reused the sliced features from the previous slicing level. All slicing configurations produce the same explanations, ensuring correctness and minimality. Performance was evaluated by summing explanation times (in seconds) over a fixed set of randomly selected instances. For each slicing level, we also measured the number of binary variables removed due to the input domain and subdomains, averaging across subdomains. These simplifications are applied to the constraints in Equations~(\ref{eq:indicator1})-(\ref{eq:z_or}), as illustrated in Example~\ref{ex:slicing_concrete}. The extent of binary variable elimination may influence the computational efficiency of our approach. As our method guarantees perfect fidelity to the underlying model by construction, we do not evaluate this metric in our experiments.

% Understanding the degree of binary variable elimination provides insight into cases where our approach yields better computational efficiency.
% This count provides insight into how slicing affects the formulation of the explanation problem and its potential impact on computational efficiency."

%For each dataset, we trained three NNs: one with two hidden layers of 16 neurons each, another with three hidden layers of the same size, and a third with four hidden layers, also with 16 neurons per layer. 

%For each, we trained three NNs with 16 neurons per layer: two hidden layers, three hidden layers, and four hidden layers.

We used five datasets from the UCI Machine Learning Repository\footnote{\url{https://archive.ics.uci.edu/ml/}} and the Penn Machine Learning Benchmarks\footnote{\url{https://github.com/EpistasisLab/penn-ml-benchmarks/}}: Auto (25 features), Hepatitis (19 features), Australian (14 features), Heart-Statlog (13 features), and Glass (9 features). For each, we trained three NNs, each with 16 neurons per layer: one with two hidden layers, one with three hidden layers, and one with four hidden layers. The NNs were implemented with TensorFlow, using a 90\%-10\% data split for training and testing. Training used a batch size of 4 and 100 epochs. Explanations were performed on the test set. For the entailment checks in Algorithm~\ref{algorithm1}, we used IBM-CPLEX\footnote{\url{https://www.ibm.com/docs/en/icos/22.1.2}} via the DOcplex API to handle MILP constraints. All experiments were run on an Intel Core i5-7200U (2.50 GHz) with 8 GB of RAM. Full details are available in our repository\footnote{\url{https://github.com/Luizfernandopq/Explications-ANNs}}.

%Explanations were performed on the test set (10% of the data), selected consistently across all slicing variations, enabling a fair comparison of average explanation times. The number of instances explained for each dataset was 21, 16, 70, 28, and 21 for Auto, Hepatitis, Australian, Heart-Statlog, and Glass, respectively. For entailment checks, we used IBM-CPLEX via the DOcplex API to handle MILP constraints. All experiments were run on an Intel Core i5-7200U (2.50 GHz) with 8 GB of RAM. Full details are available in our repository.

%used consistently across all slicing variations. This setup allowed a fair comparison of average explanation times across different slicing levels. 
%The number of instances explained for each dataset was 21, 16, 70, 28, and 21 for Auto, Hepatitis, Australian, Heart-Statlog, and Glass, respectively. 

%\textcolor{red}{COLOCAR CONFIGURAÇÔES DO COMPUTADOR QUE EXECUTOU OS EXPERIMENTOS.}
%https://github.com/Luizfernandopq/Explications-ANNs

%For instance, in the Auto dataset, the explanation time increased from 12.56 (no slicing) to 13.80 (three slices), and in the Hepatitis dataset, from 4.49 (no slicing) to 9.97 (three slices).

The results for 2-, 3-, and 4-layer NNs are shown in Tables~\ref{table:2-layers}–\ref{table:4-layers}, respectively. The number of features of each dataset is indicated in parentheses. For 2-layer NNs (Table~\ref{table:2-layers}), slicing did not improve explanation time and often increased it, with little to no binary variable removal. This suggests that, for smaller architectures, the overhead of slicing outweighs its benefits, and performance gains may only appear beyond a certain NN size.

\begin{table}[h]
\centering
\scriptsize
\caption{Explanation time (in seconds) and average \% binary variables removed by number of slices for 2-layers NNs.}
\label{table:2-layers}
\begin{tabular}{|l|c|c|c|c|c|}
\hline
\textbf{Dataset} & \textbf{Metric} & \textbf{0 slices} & \textbf{1 slice} & \textbf{2 slices} & \textbf{3 slices} \\
\hline
Auto & Exp Time (s) & 12.56 & 13.17 & 13.32 & 13.80 \\
 (25)          & \% Bin Rem & 0.00 & 0.00 & 0.00 & 0.00 \\
\hline
Hepatitis & Exp Time (s) & 4.49 & 5.83 & 6.74 & 9.97 \\
 (19)               & \% Bin Rem & 0.00 & 0.00 & 0.00 & 0.00 \\
\hline
Australian  & Exp Time (s) & 32.20 & 33.25 & 33.01 & 34.09 \\
(14)               & \% Bin Rem & 0.00 & 1.00 & 1.50 & 2.13 \\
\hline
Heart-Statlog & Exp Time (s) & 9.65 & 9.85 & 10.75 & 13.12 \\
 (13)                   & \% Bin Rem & 0.00 & 0.00 & 1.00 & 1.38 \\
\hline
Glass  & Exp Time (s) & 5.80 & 5.97 & 6.60 & 6.87 \\
(9)          & \% Bin Rem & 0.00 & 0.00 & 0.00 & 0.13 \\
\hline
\end{tabular}
\end{table}

%The Australian dataset shows a slight reduction of 5.00\% with two slices. 

%The Heart-Statlog dataset demonstrates the most consistent improvements, with an 11.22\% reduction for two slices and 15.67\% for three slices.

%Next, we discuss the results for 3-layers NNs in Table~\ref{table:3-layers}. The experimental results indicate that introducing slices can, in certain cases, improve explanation times. Moreover, the number of binary variables removed is more noticeable. 
For 3-layer NNs (Table~\ref{table:3-layers}), slicing sometimes improved explanation time, especially when accompanied by notable binary variable removal. In the Australian dataset, binary variable removal reached 16\% with three slices, and the best explanation time reduction (5.00\%) occurred with two slices (12.75\% removal). Heart-Statlog followed a similar pattern, with up to 3.50\% removal and significant time reductions of 15.67\% (three slices) and 11.22\% (two slices). In contrast, Hepatitis and Glass showed low binary variable removal and increasing explanation times with more slices, indicating that slicing introduced overhead without sufficient simplification. The Auto dataset showed an interesting behavior: no binary variables were removed at any slicing level, yet explanation time improved (13.40\% with one slice, 9.36\% with three). This suggests that efficiency gains can also stem from other structural changes induced by slicing, not just from binary variable elimination.

\begin{table}[h]
\centering
\scriptsize
\caption{Explanation time (in seconds) and average \% of binary variables removed by number of slices for 3-layers NNs. The values in bold represent the slices that achieved a reduction in explanation time compared to 0 slices.}
\label{table:3-layers}
\begin{tabular}{|l|c|c|c|c|c|}
\hline
\textbf{Dataset} & \textbf{Metric} & \textbf{0 slices} & \textbf{1 slice} & \textbf{2 slices} & \textbf{3 slices} \\
\hline
Auto     & Exp Time (s) & 56.40 & \textbf{48.84} & 68.86 & \textbf{51.12} \\
  (25)             & \% Bin Rem & 0.00 & 0.00 & 0.00 & 0.00 \\
\hline
Hepatitis & Exp Time (s) & 15.85 & 19.82 & 19.90 & 18.59 \\
 (19)               & \% Bin Rem & 0.00 & 0.00 & 0.00 & 0.38 \\
\hline
Australian & Exp Time (s) & 149.92 & 154.09 & \textbf{142.43} & 153.18 \\
 (14)                & \% Bin Rem & 0.00 & 11.00 & 12.75 & 16.00 \\
\hline
Heart-Statlog & Exp Time (s) & 69.50 & 73.63 & \textbf{61.70} & \textbf{58.61} \\
 (13)                   & \% Bin Rem & 0.00 & 1.00 & 1.50 & 3.50 \\
\hline
Glass           & Exp Time (s) & 38.99 & 53.65 & 42.44 & 49.58 \\
(9)                   & \% Bin Rem & 1.00 & 1.00 & 1.00 & 1.13 \\
\hline
\end{tabular}
\end{table}

%Next, we discuss the results for 4-layers NNs from Table~\ref{table:4-layers}. The Australian and Heart-Statlog datasets show the most substantial benefits from slicing. In the Australian dataset, explanation times improve consistently as more slices are added, culminating in a remarkable 37.20\% reduction with three slices. 

Next, we discuss the results for 4-layers NNs from Table~\ref{table:4-layers}. The Australian and Heart-Statlog datasets show the most substantial benefits from slicing. In the Australian dataset, explanation times improve consistently as more slices are added, culminating in a remarkable 37.20\% reduction with three slices. This aligns with the progressive increase in binary variable removal, which reaches 23.38\% for three slices. Similarly, the Heart-Statlog dataset benefits across all slicing configurations, with explanation time reductions exceeding 25\% for one slice and reaching 40.46\% with three slices. These results highlight the potential of slicing to enhance efficiency in explanation tasks, especially for larger NNs. Conversely, the Auto dataset presents a case where slicing consistently increases explanation time, and no binary variables are removed at any slicing level. This reinforces the observation that, for this dataset, slicing introduces computational overhead without providing meaningful simplifications. Similarly, the Hepatitis and Glass datasets exhibit low binary variable removal (at most 1.50\% and 2.00\%, respectively), and their explanation time reductions are inconsistent. The Hepatitis dataset benefits slightly from one slice (6.28\% reduction) but deteriorates with additional slices. The Glass dataset shows a notable improvement only with three slices (21.39\% reduction).%, which may indicate that slicing effects are more dataset-dependent in shallower networks with fewer binary constraints.

%The pattern of increased removal with faster explanation times suggests that even modest reductions in binary constraints can contribute to efficiency gains.

%Similarly, the Heart-Statlog dataset benefits across all slicing configurations, with explanation time reductions exceeding 25\% for one slice and reaching 40.46\% with three slices. These results highlight the potential of slicing to enhance efficiency in explanation tasks, especially for larger NNs. 

%The Hepatitis dataset demonstrates a slight reduction of 6.28\% with one slice. The Glass dataset presents a reduction of 21.39\% with three slices.

% \begin{table}[h]
% \centering
% \caption{Explanation time by number of slices for 4-layers NNs. The values in bold represent the slices that achieved a reduction in explanation time compared to 0 slices.}
% \label{table:4-layers}
% \begin{tabular}{|c|c|c|c|c|}
% \hline
% \textbf{Dataset} & \textbf{0 slices} & \textbf{1 slice} & \textbf{2 slices} & \textbf{3 slices} \\
% \hline
% Auto (25)     & 387.34 & 454.78 & 466.26 & 460.04 \\
% \hline
% Hepatitis (19) & 463.45 & \textbf{434.36} & 510.26 & 625.97 \\
% \hline
% Australian (14) & 6156.11 & \textbf{6017.00} & \textbf{5195.69} & \textbf{3867.26} \\
% \hline
% Heart-Statlog (13) & 3317.58 & \textbf{2457.75} & \textbf{2523.63} & \textbf{1974.47} \\
% \hline
% Glass (9)          & 217.89 & 253.87 & 233.12 & \textbf{171.31} \\
% \hline
% \end{tabular}
% \end{table}

\begin{table}[h]
\centering
\scriptsize
\caption{Explanation time (in seconds) and average \% of binary variables removed by number of slices for 4-layers NNs. The values in bold represent the slices that achieved a reduction in explanation time compared to 0 slices.}
\label{table:4-layers}
\begin{tabular}{|l|c|c|c|c|c|}
\hline
\textbf{Dataset} & \textbf{Metric} & \textbf{0 slices} & \textbf{1 slice} & \textbf{2 slices} & \textbf{3 slices} \\
\hline
Auto     & Exp Time (s) & 387.34 & 454.78 & 466.26 & 460.04 \\
   (25)            & \scriptsize{\% Bin Rem} & 0.00 & 0.00 & 0.00 & 0.00 \\
\hline
Hepatitis & Exp Time (s) & 463.45 & \textbf{434.36} & 510.26 & 625.97 \\
 (19)               & \scriptsize{\% Bin Rem} & 1.00 & 1.00 & 1.50 & 1.50 \\
\hline
Australian  & Exp Time (s) & 6156.11 & \textbf{6017.00} & \textbf{5195.69} & \textbf{3867.26} \\
(14)                & \scriptsize{\% Bin Rem} & 1.00 & 15.50 & 19.00 & 23.38 \\
\hline
Heart-Statlog & Exp Time (s) & 3317.58 & \textbf{2457.75} & \textbf{2523.63} & \textbf{1974.47} \\
 (13)                   & \scriptsize{\% Bin Rem} & 3.00 & 3.00 & 3.50 & 4.75 \\
\hline
Glass           & Exp Time (s) & 217.89 & 253.87 & 233.12 & \textbf{171.31} \\
 (9)                  & \scriptsize{\% Bin Rem} & 1.00 & 1.00 & 1.50 & 2.00 \\
\hline
\end{tabular}
\end{table}

%Although additional slices do not further improve performance, this result reinforces the idea that the effectiveness of slicing may depend on selecting the appropriate number of slices for a given network and dataset.
%, suggesting that in some cases, even a small amount of slicing can provide efficiency gains.

%Although some improvements are relatively modest, these results highlight that even slicing a single attribute can reduce explanation time. Additionally, using multiple slices often improves computational efficiency, likely due to the reduced complexity of processing smaller input domains. While the impact of slicing varies across datasets, our findings provide clear evidence that it can effectively accelerate the explanation process.

Overall, the results suggest that, for deeper networks, the effectiveness of slicing is strongly influenced by the extent of binary variable removal. When this removal is substantial, as in the Australian and Heart-Statlog datasets, slicing tends to yield consistent efficiency gains. However, when the number of binary variables removed is minimal or nonexistent, slicing may introduce overhead that outweighs its potential benefits.

Additionally, the relationship between binary variable removal and explanation time improvements appears to be more dataset-dependent than strictly related to NN depth or data dimensionality. For instance, the Auto and Glass datasets both exhibit minimal binary variable removal at any slicing level. These datasets have significantly different dimensionalities (25 and 9, respectively), suggesting that the impact of slicing is not solely determined by the number of input features, but rather by dataset-specific characteristics or features selected for slicing. Certain features may better capture meaningful structure in the data, leading to greater simplification and more efficient explanations.

% Reviewer 2 (3)

% \textcolor{red}{Additionally, the relationship between binary variable removal and explanation time improvements appears to be more dataset-dependent than strictly related to NN depth or data dimensionality. This suggests that the specific feature selected for slicing may play a significant role in the overall performance of the approach, as certain features may better capture meaningful structure in the data and lead to more efficient explanations.}

\section{Conclusions}

In this paper, we propose using domain slicing to enhance the scalability of logic-based XAI for NNs. Our approach partitions the input domain into smaller subdomains, allowing simplification of logical constraints. Experiments show that domain slicing can reduce explanation time by over 40\%. Notably, the benefits of slicing are more pronounced in cases where binary variable removal becomes more effective. This improvement suggests that modest slicing can enhance efficiency without compromising correctness.

%While current logic-based methods ensure correctness, their scalability remains limited.  

%Experimental results demonstrate that domain slicing can significantly reduce explanation times—by over 40% in some configurations—particularly for deeper networks. Notably, the benefits of slicing are more pronounced in 3- and 4-layer NNs, where the simplification of constraints through binary variable removal becomes more effective. In contrast, for shallow networks, the overhead introduced by slicing can outweigh its benefits.

Moreover, the relationship between explanation time improvements and binary variable removal appears to depend more on dataset-specific characteristics than on NN depth or input dimensionality. This also suggests that the features selected for slicing can play a key role, highlighting the importance of feature selection in the slicing process. Future work may explore slicing strategies that adapt to dataset characteristics and network properties, as well as techniques for automatically identifying the most effective features for slicing.

% Reviewer 1 (4)
% textcolor{red}{
% In this paper, we propose using domain slicing to enhance the scalability of logic-based XAI for NNs. While current logic-based methods ensure correctness, their scalability remains limited. Experiments show that domain slicing can reduce explanation time by over 40\%. This improvement indicates that when combined with feature engineering, it can serve as a robust alternative. Future research may focus on devising strategies to identify features most suitable for slicing, tailored to the data characteristics and neural network properties.
% }

\begin{credits}
\subsubsection*{\ackname}
This work was partially funded by Coordenação de Aperfeiçoamento de Pessoal de Nível Superior (CAPES).%The authors acknowledge the support of the Federal Institute of Education, Science and Technology of Ceará (IFCE) through the research grant calls PIBIC No. 11/2024, issued by the PRPI/IFCE, as well as the support of Fundação Cearense de Apoio ao Desenvolvimento Científico e Tecnológico (FUNCAP) and Conselho Nacional de Desenvolvimento Científico e Tecnológico (CNPq) in the development of this work.

% A bold run-in heading in small font size at the end of the paper is
% used for general acknowledgments, for example: This study was funded

% \subsubsection{\discintname} ...
% It is now necessary to declare any competing interests or to specifically
% state that the authors have no competing interests. Please place the
% statement with a bold run-in heading in small font size beneath the
% (optional) acknowledgments\footnote{If EquinOCS, our proceedings submission
% system, is used, then the disclaimer can be provided directly in the system.},
% for example: The authors have no competing interests to declare that are
% relevant to the content of this article. Or: Author A has received research
% grants from Company W. Author B has received a speaker honorarium from
% Company X and owns stock in Company Y. Author C is a member of committee Z.
\end{credits}

\bibliographystyle{splncs04nat}
\bibliography{references}

@book{goodfellow:16,
    title={Deep Learning},
    author={Ian Goodfellow and Yoshua Bengio and Aaron Courville},
    publisher={MIT Press},
    year={2016}
}

@article{alghoul2018email,
  title={Email classification using artificial neural network},
  author={Alghoul, Ahmed and Al Ajrami, Sara and Al Jarousha, Ghada and Harb, Ghayda and Abu-Naser, Samy S},
  year={2018},
  publisher={IJARW},
  volume={2},
  number={11},
  journal={IJAER}
}

@article{musleh2019predicting,
  title={Predicting liver patients using artificial neural network},
  author={Musleh, Musleh M and Alajrami, Eman and Khalil, Ahmed J and Abu-Nasser, Bassem S and Barhoom, Alaa M and Naser, SS Abu},
  journal={IJAISR},
  volume={3},
  number={10},
  year={2019}
}

@inproceedings{koh2017understanding,
  title={Understanding black-box predictions via influence functions},
  author={Koh, Pang Wei and Liang, Percy},
  booktitle={34th ICML},
  year={2017}
}

@inproceedings{elboher2020abstraction,
  title={An abstraction-based framework for neural network verification},
  author={Elboher, Yizhak Yisrael and Gottschlich, Justin and Katz, Guy},
  booktitle={32nd CAV},
  year={2020}
}

@inproceedings{Ribeiro2016Lime,
  title={``{W}hy should {I} trust you?'' Explaining the predictions of any classifier},
  author={Ribeiro, Marco Tulio and Singh, Sameer and Guestrin, Carlos},
  booktitle={22nd ACM SIGKDD},
  year={2016}
}

@inproceedings{Ribeiro2018Anchors,
  title={Anchors: High-precision model-agnostic explanations},
  author={Ribeiro, Marco Tulio and Singh, Sameer and Guestrin, Carlos},
  booktitle={32nd AAAI},
  year={2018}
}

@inproceedings{ignatiev2019abduction,
  title={Abduction-based explanations for machine learning models},
  author={Ignatiev, Alexey and Narodytska, Nina and Marques-Silva, Joao},
  booktitle={33rd AAAI},
  year={2019}
}

@inproceedings{audemard2022preferred,
  title     = {On Preferred Abductive Explanations for Decision Trees and Random Forests},
  author    = {Audemard, Gilles and Bellart, Steve and Bounia, Louenas and Koriche, Frederic and Lagniez, Jean-Marie and Marquis, Pierre},
  booktitle = {31st {IJCAI}},
  year      = {2022}
}

@inproceedings{katz,
  title={Pruning and Slicing Neural Networks using Formal Verification},
  author={Lahav, Ori and Katz, Guy},
  booktitle={FMCAD},
  year={2021}
}

@inproceedings{shi2018symbolic,
author = {Shih, Andy and Choi, Arthur and Darwiche, Adnan},
title = {A Symbolic Approach to Explaining Bayesian Network Classifiers},
year = {2018},
booktitle = {27th IJCAI},
}

@article{rocha2025generalizing,
  title={Generalizing Logic-based Explanations for Machine Learning Classifiers via Optimization},
  author={Rocha Filho, Francisco Mateus and da Rocha Neto, Ajalmar R{\^e}go and Rocha, Thiago Alves},
  journal={Expert Systems with Applications},
  volume={289},
  year={2025},
  publisher={Elsevier}
}

@article{amparore2021trust,
  title={To trust or not to trust an explanation: using {LEAF} to evaluate local linear {XAI} methods},
  author={Amparore, Elvio and Perotti, Alan and Bajardi, Paolo},
  journal={PeerJ Computer Science},
  volume={7},
  year={2021},
  publisher={PeerJ Inc.}
}

@book{kroening2016decision,
  title={Decision procedures},
  author={Kroening, Daniel and Strichman, Ofer},
  year={2016},
  publisher={Springer}
}

@inproceedings{bassan2023towards,
author="Bassan, Shahaf
and Katz, Guy",
title="Towards Formal {XAI}: Formally Approximate Minimal Explanations of Neural Networks",
booktitle="TACAS",
year="2023"
}

@inproceedings{bjorner2023formal,
  title={Formal {XAI} via syntax-guided synthesis},
  author={Bj{\o}rner, Katrine and Judson, Samuel and Cano, Filip and Goldman, Drew and Shoemaker, Nick and Piskac, Ruzica and K{\"o}nighofer, Bettina},
  booktitle={1st AISoLA},
  pages={119--137},
  year={2023}
}

@inproceedings{wang2021probabilistic,
  title     = {Probabilistic Sufficient Explanations},
  author    = {Wang, Eric and Khosravi, Pasha and Van den Broeck, Guy},
  booktitle = {30th {IJCAI}},
  year      = {2021}
}

@article{milp1971,
  title={Experiments in mixed-integer linear programming},
  author={Michel B{\'e}nichou and Jean-Michel Gauthier and P. Girodet and Gerard Hentges and Gerard Ribi{\`e}re and O. Vincent},
  journal={Mathematical Programming},
  year={1971},
  volume={1},
  pages={76-94}
}

% \end{thebibliography}
\end{document}